\documentclass[aps,10pt,amsmath]{revtex4}
\usepackage{epsfig}
\begin{document}

\title{Continuum shell model: From Ericson to conductance fluctuations}

\author{G.L.~Celardo$^{1}$, F.M.~Izrailev $^{2,3}$, S.~Sorathia$^{2}$, V.G.~Zelevinsky$^{3}$,
G.P.~Berman$^{4}$}

\address{$^{1}$Physics Department, Tulane University, New Orleans, LA 70118, USA}

\address{$^{2}$Instituto de F\'isica, Universidad Aut\'onoma de Puebla,
Apartado Postal J-48, Puebla, Pue., 72570, M\'exico}

\address{$^{3}$NSCL and Department of Physics and Astronomy, Michigan State
University, East Lansing, Michigan 48824-1321, USA}

\address{$^{4}$Theoretical Division and CNLS, Los Alamos National Laboratory, Los
Alamos, New Mexico 87545, USA}

\begin{abstract}
We discuss an approach for studying the properties of mesoscopic
systems, where discrete and continuum parts of the spectrum are
equally important. The approach can be applied (i) to stable heavy
nuclei and complex atoms near the continuum threshold, (ii) to
nuclei far from the region of nuclear stability, both of the regions
being of great current interest, and (iii) to mesoscopic devices
with interacting electrons. The goal is to develop a new consistent
version of the continuum shell model that simultaneously takes into
account strong interaction between fermions and coupling to the
continuum. Main attention is paid to the formation of compound
resonances, their statistical properties, and correlations of the
cross sections. We study the Ericson fluctuations of overlapping
resonances and show that the continuum shell model nicely describes
universal properties of the conductance fluctuations.
\end{abstract}

\maketitle

\subsection{Introduction}

The focus of interest in low-energy nuclear physics has recently
concentrated on nuclei far from the region of stability. The success
of the nuclear shell model with strong interaction between nucleons
urges one to look for ways of incorporating this rich experience
into a more general context that would properly include the
continuum. The continuum shell model is expected to combine the
physics of intrinsic structure and reactions in contrast to the
traditional spectroscopy and reaction theory which were historically
developed as different branches of nuclear physics. In this
approach, the properties of open channels, threshold behavior,
symmetries, and spectroscopic factors enter the overall theory, so
that a fully consistent picture uniting structure and reaction
mechanisms is created. With great successes of mesoscopic physics
(complex atoms and molecules, micro- and nano-devices of condensed
matter, atomic and molecular traps, and prototypes of quantum
computers), the nuclear problem becomes a typical part of general
science of open mesoscopic systems.

In this paper we discuss a new approach \cite{CIZB07} for studying
collective phenomena in nuclei and quantum dots which emerge as a
result of strong interactions between fermions coupled to the
continuum. In this problem the standard shell model techniques,
adjusted for the bound states, cannot work and continuum effects
must be properly accounted for. At a certain strength of the
continuum coupling, the system of strongly interacting fermions
undergoes a collective restructuring. This is an area where the
conventional division of nuclear physics into "structure" and
"reactions" becomes inappropriate and two views of the process, from
the inside (structure and properties of bound states) and from the
outside (cross sections of reactions), have to be unified.

Conventional theory of statistical reactions does not answer the
question of interplay between reactions and internal structure
determined by the character of interactions between the
constituents. Here more detailed considerations are required based
on the generalization of the shell model of nuclear reactions
\cite{MW}. Such an extension requires statistical assumptions
concerning intrinsic dynamics and the coupling to the continuum
\cite{agassi75,verbaarschot85}. To account for specificity of the
system, one has to go beyond standard random matrix approaches
\cite{brody81,GMW98,MK04} utilizing the Gaussian Orthogonal Ensemble
(GOE). The consistent description based on the continuum shell model
\cite{MW}, as well as more phenomenological approaches
\cite{M67,simonius74}, indicates the presence of a sharp
restructuring of the system when the widths of resonances become
comparable to their energy spacings. This phenomenon carries
features of a quantum phase transition with the strength of
continuum coupling playing the role of a control parameter.

As was numerically observed in the shell model framework
\cite{kleinwachter85}, the distribution of resonance widths rapidly
changes in the transitional region in such a way that a number of
very broad resonances equal to a number of open decay channels
absorb the lion's share of the total width of all overlapped
resonances, while the remaining states become very narrow. The
corresponding theory was suggested in \cite{SZPL88,SZNPA89,SZAP92},
where the mechanism of this restructuring was understood to be
associated with the nature of the effective non-Hermitian
Hamiltonian \cite{MW} that describes the intrinsic dynamics after
eliminating the channel variables. The segregation of short-lived
broad resonances from long-lived trapped states was shown to be
similar to the superradiance \cite{dicke54} in quantum optics
induced by the coupling of atomic radiators through the common
radiation field, an analog of coherent coupling of many overlapped
intrinsic states through continuum decay channels. Later a general
character of the phenomenon was demonstrated for systems with the
GOE intrinsic dynamics and many open channels \cite{ISS94,SFT99}.
Modern versions of the shell model in continuum
\cite{rotter91,VZCSM05} are based on the effective Hamiltonian and
naturally reveal the superradiance phenomenon as an important
element. The transition to this regime should be taken into account
in all cases when a physical system is strongly coupled to the
continuum, see for example \cite{VZWNMP04} and references therein.

In the present work, using the same framework as in \cite{CIZB07},
we study the interplay between the intrinsic dynamics and
statistical properties of cross sections comparing the results with
those of conventional approaches, namely Hauser-Feshbach average
cross sections and Ericson fluctuations and correlations. We also
analyze how the universal conductance fluctuations appear in our
model, how they depend on the intrinsic chaos and what is the role
of correlations between cross sections of various processes.

\subsection{The model}

We consider a generic system of $n$ interacting fermions distributed
over $m$ single-particle levels of energies $\epsilon_s$
(``mean-field orbitals"). The fermions are coupled to continuum
through the finite number $M$ of open decay channels. In the
many-body representation, the Hamiltonian matrix of size
$N=m!/[n!(m-n)!]$ entirely determines the intrinsic dynamics of the
closed system, and our main interest is to reveal how the properties
of the reactions depend on the degree of intrinsic chaos that is due
to the inter-particle interaction. According to the well developed
formalism \cite{MW,SZNPA89,VZWNMP04}, the behavior of an open system
can be described in terms of the effective non-Hermitian Hamiltonian
consisting of two terms,
\begin{equation}
{\cal H}= H - \frac{i}{2}\, W\,;\,\,\,\,\,\,\,\,\,\,\, W_{ij}=\sum
_{c=1}^M A_{i}^{c}A_j^c.                         \label{1}
\end{equation}
Here $H$ is a Hermitian matrix that can be also influenced by the
presence of the continuum \cite{rotter91,VZCSM05,CSM06}, and $W$
stands for the coupling between intrinsic many-body states labeled
as $i,j,...$, and open decay channels $a,b,c...$. The factorized
structure of $W$ is dictated by the requirement of the unitarity of
the scattering matrix. In what follows we assume that the transition
amplitudes $A_{i}^{c}$ between intrinsic states $|i\rangle$ and
channels $c$ are real quantities, therefore, we restrict ourself by
the time-invariant systems. As a result, both $H$ and $W$ are real
symmetric matrices.

The Hermitian part, $H=H_0+V$, of the full Hamiltonian (\ref{1}) is
modeled by the so-called two-body random interaction (TBRI) (see,
for example, \cite{FI97} and references therein),
\begin{equation}
\label{H}H=H_0+V; \:\:\:\:\:\:\:\:\:\:\:\:\:\:\:\:\:\:\: H_0 =
\sum_{s=1}^{m} \epsilon_{s}\,a_{s}^{\dagger }a_{s};
\:\:\:\:\:\:\:\:\:\:\:\:\:\:\:\:\:\:\: V =\frac 12\sum V_{s_1 s_2
s_3 s_4}\,a_{s_1}^{\dagger}a_{s_2}^{\dagger} a_{s_3}a_{s_4} .
\end{equation}
Here the mean-field part $H_0$ describes non-interacting particles
(or quasi-particles), and the interaction between particles is given
by $V$. In the many-particle basis $|k\rangle$ the matrix $H$ is
constructed by the Slater determinant, $\left| k\right\rangle
=a_{s_1}^{\dagger }\,.\,\,.\,\,.\,a_{s_{n}}^{\dagger }\left|
0\right\rangle$, where $a_{s}^{\dagger }$ and  $ a_{s}$ are the
creation and annihilation operators. The interaction between the
particles is considered to be of two-body nature, therefore, each
many-body matrix element $V_{lk}=\left<l|V|k\right>$ is a sum of a
number of two-body matrix elements $V_{s_1 s_2 s_3 s_4}$ involving
at most four single-particle states $|s\rangle$. In our simulations
we take $n=6,\,m=12$ that provides a sufficiently large dimension
$N=924$.

The {\sl single-particle} energies, $\epsilon_s$, are assumed to
have a Poissonian distribution of spacings, with the mean level
density $1/d_0$. The interaction $V$ is characterized by the
variance of the normally distributed two-body random matrix
elements, $\langle V_{s_1,s_2;s_3,s_4}^2\rangle=v_0^2$. Without the
interaction, $v_0=0$, the many-body states have also the Poissonian
spacing distribution $P(s)$. In the opposite limit of an extremely
strong interaction, $\lambda\equiv v_0/d_{0} \rightarrow \infty$
(or, equivalently, for $d_0=0$), the function $P(s)$ is close to the
Wigner-Dyson (WD) distribution that is typical for completely
chaotic systems \cite{brody81}. The critical value of the
interaction for the onset of strong chaos was obtained in Ref.
\cite{FI97}, it is estimated as
\begin{equation}
\label{estimate} \lambda_{{\rm cr}}=\frac{v_{{\rm cr}}}{d_0} \approx
\frac{2(m-n)}{N_s},
\end{equation}
where $N_s=n(m-n)+ n(n-1)(m-n)(m-n-1)/4$ is the number of directly
coupled many-body states in any row of the matrix $H_{ij}$. One can
expect that for a very strong interaction, $\lambda \gg
\lambda_{{\rm cr}}$, some of properties of the system may be close
to those obtained with $H$ taken from the Gaussian Orthogonal
Ensemble (GOE). For this reason, we have also considered, for
comparison, the GOE matrices in place of $H$.

In our study the amplitudes $A_i^c$ are assumed to be random
independent Gaussian variables with the zero mean and variance
\begin{equation}
\langle A_i^c A^{c'}_j\rangle=\delta_{ij}\delta^{cc'}\,\frac{
\gamma^{c}}{N}.                                \label{2}
\end{equation}
This is compatible with the GOE or TBRI models where generic
many-body states coupled to continuum have a very complicated
structure, while the decay probes specific simple components of
these states related to few open channels. Even for a weak intrinsic
interaction, one needs to have in mind that in reality the states
$|i\rangle$ have certain values of exact constants of motion, such
as angular momentum and isospin in the nuclear case. Therefore, at
sufficiently large dimension $N$, these states acquire {\sl
geometric chaoticity} \cite{ZV04} due to the almost random coupling
of individual spins. Thus, the assumption of a random nature of the
decay amplitudes is reasonable.

The parameters $\gamma^{c}$ characterize the total coupling of all
states to the channel $c$. The normalization used in Eq. (\ref{2})
is convenient if the energy interval $ND$ covered by decaying states
is finite. Here $D$ is the distance between the many-body states in
the middle of the spectrum, $D=1/\rho(0)$, where $\rho(E)$ is the
many-body level density, and $E=0$ corresponds to the center of the
spectrum. Below we neglect a possible energy dependence of the
amplitudes that is important near thresholds and is taken into
account in realistic shell model calculations \cite{VZCSM05,CSM06}.
The ratio $\gamma^{c}/ND$ characterizes the degree of overlap of the
resonances in the channel $c$. The control parameter determining the
strength of the coupling to the continuum can be written as follows,
\begin{equation} \kappa^{c}=\frac{\pi\gamma^{c}}{2ND}. \label{3a}
\end{equation}
The transition from separated to strongly overlapped resonances
corresponds to $\kappa^{c}\approx 1$. In order to keep the coupling
to continuum fixed, in our numerical calculations we renormalize the
absolute magnitude of the widths, $\gamma^{c}$, by varying the
intrinsic interaction and, therefore, the level density $\rho$.

The effective Hamiltonian allows one to construct the scattering
matrix,
\begin{equation}
S^{ab}=\delta^{ab}-i{\cal T}^{ab},                 \label{12}
\end{equation}
where
\begin{equation}
{\cal T}^{ab}(E)=\sum_{i,j}^N A_i^a\left(\frac{1}{E-{\cal H}}
\right)_{ij} A_j^b                        \label{5}
\end{equation}
are the scattering amplitudes determining the cross sections
$\sigma^{ab}(E)$ of reactions,
\begin{equation}
\sigma^{ab}(E) = |{\cal T}^{ab}(E)|^2.            \label{4}
\end{equation}
In our notations the cross sections are dimensionless since we omit
the common factor $\pi/k^{2}$. In what follows we study both the
elastic, $b=a$, and inelastic, $b \neq a$, cross sections. Note that
we ignore the smooth potential phases that are irrelevant for our
purposes.

\subsection{Transition to superradiance}

The complex eigenvalues ${\cal E}_{r}=\omega_{r}-(i/2)\Gamma_{r}$ of
${\cal H}$ coincide with the poles of the $S$-matrix and, for small
$\gamma^c$, determine energies and widths of separated resonances.
In the simulations we consider a real energy interval at the center
of the spectrum of ${\cal H}$ with the constant many-body level
density $\rho(0)=D^{-1}$. The transmission coefficient in the
channel $c$,
\begin{equation}
T^{c}=1-|\langle S^{cc}\rangle|^2=
\frac{4\kappa^{c}}{(1+\kappa^{c})^{2}},               \label{3b}
\end{equation}
is maximal (equal to 1) at the critical point, $\kappa^{c}=1$, and,
for simplicity, we assume $M$ equiprobable channels,
$\kappa^{c}=\kappa$. For each value of $\kappa$ we have used a large
number of realizations of the Hamiltonian matrices, with further
averaging over energy.

\begin{figure}[h!]
\includegraphics[width=6cm,angle=-90,viewport=150 360 480 440]{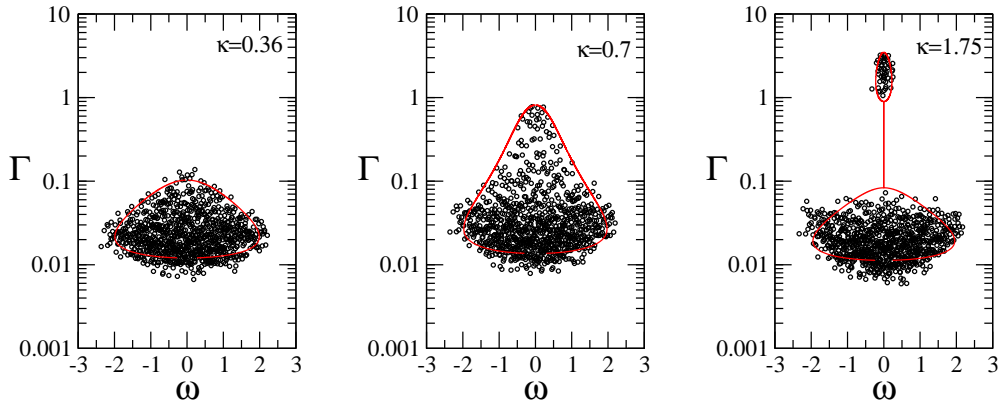}
\caption{Distribution of widths, $\Gamma$, and positions, $\omega$,
of resonances (poles of the scattering matrix) as a function of
$\kappa$. The results are shown for $n=6$ particles, $m=12$
orbitals, 50 open channels, and for a relatively strong interaction
between particles, $v_0/d_0 = 1/10$, above the strong chaos
threshold. The smooth curves are the boundaries due to the
analytical results obtained in Ref.\cite{haake} for the GOE matrices
in place of $H$. }
\end{figure}

As $\gamma^c$ grows, the resonances start to overlap, leading to a
segregation of the resonance widths
\cite{kleinwachter85,SZPL88,SZNPA89,haake} occurring at $\kappa
\approx 1 $, see Fig.~1. The widths of $M$ resonances are increasing
at the expense of the remaining $N-M$ resonances. For a weak
coupling, $\kappa \ll 1 $, the widths are given by diagonal matrix
elements, $\Gamma_{i}= \langle i|W |i \rangle= \sum_{c=1}^M
(A_i^c)^2$, and the mean width is $\langle\Gamma\rangle= \gamma
M/N$. In the limit of strong coupling, $\kappa \gg 1 $, the widths
of $M$ broad resonances converge to the non-zero eigenvalues of the
matrix $W$ that has a rank $M$ due to its factorized structure. As
for the remaining ``trapped" $(N-M)$ states, their widths decrease
as $1/\gamma$. From our data one can conclude that the transition to
the superradiance in dependence on the control parameter $\kappa$
turns out to be quite sharp, and may be treated as a kind of the
phase transition.

\begin{figure}[h!]
\includegraphics[width=4cm,angle=-90,viewport=70 100 520 440]{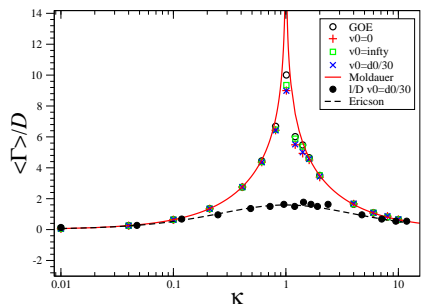}
\caption{Normalized average width versus $\kappa$ for $n=6, m=12$
and $M=10$. The average was performed over the resonances around the
center of the energy spectrum. Open circles refer to the GOE case,
pluses to $\lambda=0$, squares to $\lambda \rightarrow\infty$, and
crosses to $\lambda=1/30$. Solid curve shows the MS-expression
(\ref{mold}). Full circles stand for the normalized correlation
length at $\lambda=1/30$, see in the text. The dashed curve shows
the Weisskopf relation (\ref{62}). }
\end{figure}

In Fig.~2 we show how the {\sl average width} (normalized to the
mean level spacing $D$) depends on $\kappa$ for different values of
the interaction $\lambda$. The figure demonstrates a sharp change in
the width distribution at the transition point $\kappa =1$. The
results are compared with the expression
\begin{equation}
\frac{\langle\Gamma\rangle}{D}= \frac {M}{\pi} \ln
\left|\frac{1+\kappa}{1-\kappa}\right|             \label{mold}
\end{equation}
that is obtained from the Moldauer-Simonius (MS) relation
\cite{M67,S74} for $M$ equivalent channels. We have extrapolated the
MS-relation to the region with $\kappa > 1$ for which only $N-M$
narrow resonances were taken into account. For one channel in the
GOE case this result was obtained in
Refs.~\cite{SZPL88,SZNPA89,ISS94}; here we see that the
MS-expression is also valid for a large number of channels. Note
that the results are independent of the degree of intrinsic
chaoticity; this fact was explained in Ref.\cite{CIZB07}. According
to \cite{SFT99}, the divergence of $\langle\Gamma\rangle$ at
$\kappa=1$ is due to the (non-integrable) power-law behavior for
large $\Gamma$, see below; in the numerical simulation, see Fig.~2,
there is no divergence because of the finite number of resonances,
although the trend is clearly seen.

As for the width distribution $P(\Gamma)$, in Ref.~\cite{CIZB07} it
was shown that for a weak coupling, $\kappa \ll 1$, the conventional
$\chi^2_{M}$-distribution is valid for any strength of the
interaction. However, when $\kappa$ grows, a clear dependence on the
interaction strength emerges. As first noted in Ref.~\cite{Mol68},
as $\kappa$ increases, $P(\Gamma)$ becomes broader than the
$\chi^2_{M}$ distribution. For large $\kappa$, both for the GOE and
for $\lambda\rightarrow \infty$, $P(\Gamma)$ is again given by the
$\chi^2_M$ distribution, contrary to the cases of the finite
interaction strength. Our results are in a good agreement with the
analytical predictions \cite{mizutori93,SFT99} obtained for the GOE
case (neglecting the emergence of broad resonances in the regime of
strong overlap at $\kappa > 1$). In particular, for large number of
channels, the tails of $P(\Gamma)$ decrease as $\Gamma ^{-2}$, thus
leading to the divergence of $\langle\Gamma\rangle$ at $\kappa=1$.

\subsection{Ericson fluctuations}

According to the conventional theory of Ericson fluctuations
\cite{ericson63,brink63,EMK66}, the scattering amplitude  can be
written as the sum of the average and fluctuating parts, ${\cal
T}^{ab}(E)=\langle{\cal T}^{ab}(E)\rangle + {\cal T}^{ab}_{{\rm
fl}}(E)$, with $ \langle{\cal T}^{ab}_{{\rm fl}}(E)\rangle =0$.
Correspondingly, the average cross section, $\sigma= |{\cal T}|^2$,
can be also divided into two contributions, $
\langle\sigma\rangle=\langle\sigma_{{\rm dir}}\rangle + \langle
\sigma_{{\rm fl}}\rangle$. Here the direct reaction cross section,
$\langle \sigma_{{\rm dir}} \rangle$, is determined by the average
scattering amplitude only, while $\langle \sigma_{{\rm fl}} \rangle$
is the fluctuational cross section (also known as the compound
nucleus cross section).

The main assumption of the standard theory is that in the regime of
overlapping resonances, $\langle\Gamma\rangle > D$, the fluctuating
amplitudes can be written as ${\cal T}^{ab}_{{\rm fl}}(E) = \xi +i
\eta $ where both $\xi$ and $\eta$ are Gaussian random variables
with zero mean. This is supported by the fact that for $\langle
\Gamma \rangle \gg D$ both $\xi$ and $\eta$ are the sums of a large
number of random variables. This assumption can be associated with
the other one, namely, with statistical independence of poles
(resonance energies) and residues (resonance amplitudes). As a
result, the statistical methods can be developed that lead to
various conclusions. In the frame of our model we have tested the
most important of the Ericson predictions concerning the resonance
widths and cross sections (see also Refs.\cite{CIZB07}), paying
special attention to the dependence on the intrinsic interaction
strength, $\lambda$.

\paragraph{Fluctuations of resonance widths}

In the theory of Ericson fluctuations it is usually assumed that the
fluctuations of resonance widths are small for a large number of
channels, $w(\Gamma) /\langle\Gamma\rangle^2 \ll 1$, where
$w(\Gamma)$ stands for the variance of widths. In justification of
this statement \cite{ericson63}  it is argued that in the
overlapping regime the width of a resonance can be presented as a
sum of $M$ partial widths. Assuming that they obey the Porter-Thomas
distribution, the total width is expected to have a $\chi^2_M$
distribution, so that $w (\Gamma)/\langle\Gamma\rangle^2 =2/M$ is
small for $M\gg 1$.

However, recently we have found \cite{CIZB07} that for large values
of $\kappa$ the distribution of the widths strongly differs from the
$\chi^2_M$ distribution. The data show that, as $\kappa$ increases,
the normalized variance, $w(\Gamma)/\langle\Gamma\rangle^2$, also
increases, remaining very large even for $M=20\gg 1$. Thus, contrary
to the traditional belief, the relative variance of widths does not
become small for large number of channels. The data confirm that for
non-overlapping resonances (small $\kappa \ll 1$) there is no
dependence on $\lambda$ and $w(\Gamma)/\langle\Gamma\rangle^2$
decreases as $2/M$ for all the ensembles, as expected. However, as
$\kappa$ grows, the dependence on $\lambda$ emerges: the weaker the
intrinsic chaos (and, consequently, the more ordered is the
intrinsic spectrum) the larger are the width fluctuations. For large
$\kappa \gg 1$ and strong interaction between particles, $\lambda
> \lambda_{cr}$, the width distribution is again given by the
$\chi^2_M$ distribution, contrary to the cases of the finite
interaction strength.

\paragraph{Cross sections}

Speaking about cross sections and their fluctuations, it is
convenient to use the so-called {\sl elastic enhancement factor},
that for $M$ equivalent channels takes the form,
\begin{equation}
F=\frac{\langle \sigma_{{\rm fl}}^{aa}\rangle}{\langle \sigma_{{\rm
fl}}^{ab} \rangle}, \label{43}
\end{equation}
where $b\neq a$. Here $\langle\sigma_{{\rm fl}}^{ab}\rangle$ stands
for the fluctuational part of cross sections (elastic for $a=b$ and
inelastic for $a\neq b$). In the case of equal channels, we obtain
\begin{equation}
\langle \sigma_{{\rm fl}}^{ab} \rangle=\frac{1-|\langle
S^{aa}\rangle|^2}{F+M-1}=\frac{T}{F+M-1},       \label{45}
\end{equation}
therefore, $\langle \sigma_{{\rm fl}}^{aa} \rangle=F\langle
\sigma_{{\rm fl}}^{ab} \rangle$, where $T$ is the transmission
coefficient, see Eq.~(\ref{3b}). Since the transmission coefficient
$T$ does not depend on $\lambda$, the only dependence on $\lambda$
in Eq.~(\ref{45}) is contained in the elastic enhancement factor
$F$. The same seems to be correct even when the channels are
non-equivalent, according to the results of Ref.~\cite{muller87}. It
should be pointed out that $F$ also depends on $\kappa$.
Specifically, with an increase of $\kappa$ from zero, the value of
$F$ decreases, being confined by the interval between $3$ and $2$,
see discussion in Ref.~\cite{CIZB07}.

As one can see from Eq.(\ref{45}), with an increase of the number of
channels the dependence on the interaction strength $\lambda$
disappears for the fluctuational {\it inelastic} cross section. In
contrast, the fluctuational {\it elastic} cross section manifests a
clear dependence on $\lambda$, according to the data of
Ref.~\cite{CIZB07}. This fact allows one to directly relate the
value of the enhancement factor $F$ to the strength $\lambda$ of
interaction between the particles. Specifically, the more regular is
the intrinsic motion, the higher is the average cross section. For a
large number of channels, the $\lambda$-dependence of the elastic
cross section is in agreement with the estimate, $\langle
\sigma_{{\rm fl}}^{aa} \rangle \rightarrow FT/M$. \ \\

\begin{figure}[h!]
\includegraphics[width=4.5cm,viewport=30 60 460 480]{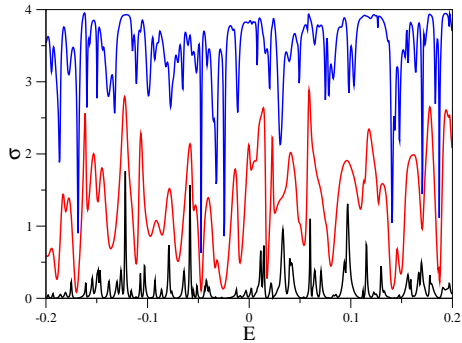}
\caption{Elastic cross-section versus energy $E$ for different
strength of coupling to the continuum: for weak, $\kappa=0.07$,
strong, $\kappa=0.7$, and very strong, $\kappa=7.0$, for $M=10$.
Other parameters are the same as in Fig.1.}
\end{figure}

\paragraph{Fluctuations of cross sections}

The fluctuations of both elastic and inelastic cross sections
strongly depend on the coupling to the continuum, see examples in
Fig.~3. According to the standard Ericson theory, in the region of
strongly overlapping resonances, $\kappa \approx 1$, the variance of
fluctuations of both elastic and inelastic cross sections can be
expressed via the average cross sections. Specifically, for equal
channels one gets,
\begin{equation}
{\rm Var}(\sigma)=\langle\sigma_{{\rm fl}}\rangle \left( 2 \langle
\sigma_{{\rm dir}}\rangle+\langle\sigma_{{\rm fl}}\rangle
\right)=\langle\sigma_{{\rm fl}}\rangle ^2.
                                              \label{31}
\end{equation}
The latter equality is written for the absence of direct processes,
that is our case. Thus, one can expect the relation, ${\rm
Var}(\sigma)/ \langle\sigma_{{\rm fl}}\rangle ^2 = 1$, for a large
number of channels. Our data confirm this relation for inelastic
cross sections, independently of the interaction strength $\lambda$,
and for $M > 10$.

As for the elastic cross sections, the expression (\ref{31}) emerges
for a quite large number of channels, $M \geq 25$. This fact of slow
convergence to the limiting value was also found analytically in
Refs. \cite{DB1,DB2} for the GOE case, any number of channels and
any coupling strength with the continuum. However, simple
expressions were derived only for $MT \gg 1$. Our data reveal a
clear dependence on $\lambda$ that is stronger for the
non-normalized variance ${\rm Var}(\sigma)$ than for the normalized
variance ${\rm Var}(\sigma)/ \langle\sigma_{{\rm fl}}\rangle ^2$.
This effect was explained in Ref. \cite{CIZB07}.

Another well known prediction of Ericson theory is that for strongly
overlapping resonances both inelastic and elastic cross sections
have the exponential distribution,
\begin{equation}
P(x)= e^{-x}, \quad x=\frac{\sigma_{{\rm fl}}}{\langle\sigma_{{\rm
fl}}\rangle}.                                 \label{54}
\end{equation}
Indeed, our data for inelastic cross sections show good agreement
with this prediction, independently of the interaction strength
$\lambda$ and for large number of channels, $M \geq 10$. For the
elastic cross sections, the exponential distribution arises for
larger number of channels, $M \geq 25$, with a weak dependence on
$\lambda$. In general, one can say that the exponential form of the
fluctuations is a good approximation in the region $\kappa \approx
1$.

\paragraph{Correlation functions}

The correlation function of cross sections is defined as
\begin{equation}
C(\epsilon)=\langle \sigma(E)\sigma(E+\epsilon)\rangle- \langle
\sigma(E) \rangle^2.                         \label{34}
\end{equation}
By assuming the Gaussian form of distribution for fluctuating
amplitudes ${\cal T}_{{\rm fl}}$, one can obtain that the
correlation function has a Lorentzian form, with the correlation
length equal to the average width,
$l_{\sigma}=\langle\Gamma\rangle$. In our model we have found that
$l_{\sigma} \approx l_{S}$ for a large number of channels in the
region $\kappa \approx 1$, and for any interaction strength
$\lambda$. Here, $l_{S}$ is the correlation length obtained from the
correlation function of the scattering matrix. On the contrary, for
smaller $M$, our data show that $l_{S}<l_{\sigma}$, and this
difference grows for the weaker interaction between the particles.

For large $M$, the correlation functions have, indeed, the
Lorentzian form for all $\lambda$. Meanwhile, for a small number of
channels the correlation function is not Lorentzian, in agreement
with the results of \cite{dittes92}. It is important to stress that
for any $M$, the correlation length is different from the average
width, apart from the region of small $\kappa$. Instead, for a large
number of channels, the correlation length is determined by the
Weisskopf relation, see \cite{lehmann95} and references therein,
\begin{equation}
\frac{l}{D}=\frac{MT}{2 \pi}=\frac{M}{2 \pi} \frac{4
\kappa}{(1+\kappa)^2}.                          \label{62}
\end{equation}

The Weisskopf relation (\ref{62}) is confirmed by our data in Fig.
2, it has been also derived in Ref. \cite{agassi75} for small values
of the ratio $m=M/N$, in the overlapping regime for the TBRI model
with the infinite interaction, as well as in Ref.
\cite{verbaarschot85} for the GOE ensemble. The fact that the
correlation length is not equal to the average width was recognized
long ago, see Refs. \cite{moldauer75} and \cite{brody81}.

\subsection{Conductance fluctuations}

Our model can be also used to study conductance and its fluctuations
for quantum dots with interacting electrons. In such an application
one can treat $M/2$ channels as the {\it left channels}
corresponding to incoming electron waves, and other $M/2$ channels,
as the {\it right channels} corresponding to outgoing waves. Then,
one can define the conductance $G$ in a standard way,
\begin{equation}
G=\sum_{a=1}^{M/2} \sum_{b=M/2 +1}^M \sigma^{ab} \label{G}
\end{equation}

\vspace{3.0cm}

\begin{figure}[h!]
\includegraphics[width=4.5cm,angle=-90,viewport=150 360 480 440]{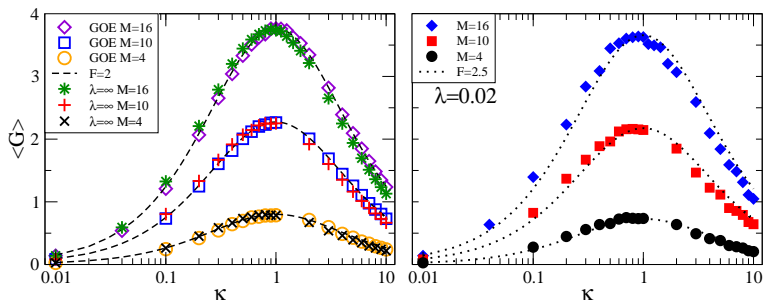}
\caption{Average conductance versus $\kappa$ for the GOE and very
strong interaction, $\lambda \rightarrow \infty$ (left panel), and
for a weak interaction, $\lambda= 0.02$ (right panel) for different
number of channels, $M=4,10,16$. Dashed curves correspond to the
analytical expression (\ref{G-av}) with $F=2.0$, (left panel), and
dotted curves with $F=2.5$, (right panel). The data are obtained for
$n=6$ particles, $m=12$ orbitals, and $M=4,10,16$ channels. }
\end{figure}

Therefore, the properties of the conductance are entirely determined
by the inelastic cross-sections, $b\neq a$. In our case of
equivalent channels the average conductance reads,
\begin{equation}\label{G-av}
\langle G \rangle =\frac{M^2}{4}\langle\sigma^{ab}\rangle=
\frac{M^2}{4}\frac{T}{F+M-1}\rightarrow \frac{MT}{4}
\end{equation}
where $T=4\kappa/(1+\kappa)^2$ and the last expression is valid for
$M \gg 1$. As one can see, the influence of the inter-particle
interaction is due to the enhancement factor $F$ only. This factor
ranges from $F=3$ for the absence of the interaction, $\lambda=0$,
to $F=2$ for a very strong interaction that can be modeled by the
GOE ensemble \cite{gorin02}. Since currently there is no theory
relating the enhancement factor to the degree of chaos in the
internal dynamics, we used this factor as the fitting parameter. In
Fig. 4 we report our numerical data obtained for the average
conductance in dependence on the coupling strength $\kappa$ for
different values of $M$. As expected, for a very strong interaction
(absence of the mean field part, $d_0=0$, or, the same,
$\lambda\rightarrow\infty$) the results are very close to those
obtained for the GOE Hamiltonian $H$, and there is a good agreement
with the expression (\ref{G}) with the theoretical value $F=2$. On
the other hand, for a relatively weak interaction, $\lambda = 0.02$,
(below the chaos threshold in a closed model) the data can be
described by the same expression with $F=2.5$ that is in the middle
of the region between the Poisson and Wigner-Dyson statistics for
the energy spacings. As one can see, for this intermediate case the
correspondence between the relation (\ref{G-av}) and data is also
good. Note, that for a very large number of channels the influence
of the internal dynamics disappears.

Our main interest was in the variance, ${\rm Var}(G)$, of the
conductance when we change the energy $E$ around the center of the
energy spectrum. In Fig.~5 we report the data for the variance for
the GOE case (left part) and for a relatively weak interaction
$\lambda =0.02$ in dependence on the coupling parameter $\kappa$ and
number $M$ of channels. First observation concerns a clear evidence
that for a weak coupling, $\kappa \approx 0.1$, the role of the
interaction should not be neglected. This fact is in contrast with
the average conductance for which the influence of the internal
chaos seems to be much less in comparison with the conductance
fluctuations.

\begin{figure}[h!]
\includegraphics[width=4.5cm,angle=-90,viewport=150 360 480 440]{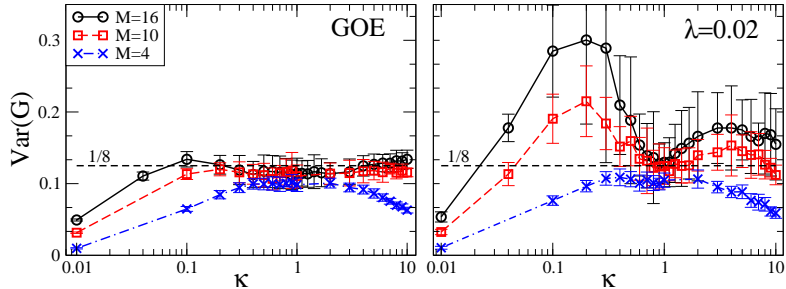}
\caption{Variance of the conductance versus $\kappa$ for the GOE
(left panel), and for $\lambda=0.02$, (right panel) . Circles,
squares and crosses stand for $M=16,10,4$ channels, respectively.
The data are given for $n=6$ and $m=12$.}
\end{figure}

Second important observation is that at the transitional point,
$\kappa \approx 1$ (known in the solid state applications as
``perfect coupling"), practically there is no influence of the
inter-particle interaction. This fact is quite instructive, it
manifests an important role of the superradiance when the (strong)
interaction between the particles is mainly due to the continuum.
Also, the dependence of fluctuations on the number of channels at
the transitional region is relatively weak (and disappears with an
increase of $M$). In the theory of conductance fluctuations it is
known that for a perfect coupling the variance takes the universal
value in the limit of a very large number of channels. This value
for time-invariant systems is either 2/15 or 1/8 depending on the
type of transport (ballistic or diffusive, both for a bulk
scattering, see, e.g. \cite{been}). It is quite instructive that our
data confirm the value 1/8 for the GOE at $\kappa=1$ with a very
good accuracy. Specifically, the data also reveal small corrections
due to the finite values of $M$, that reduce the variance to that of
slightly below 1/8, in agreement with the analytical results
obtained in Ref. \cite{been}). Also, the data indicate the value
2/15 for the TBRI model with a moderate interaction between the
particles.

We have performed a detailed analysis of the emergence of the
universal value 1/8. We should point out that one has to take into
account the correlations between different cross sections in the
expression for the variance of the conductance. Otherwise, one gets
the incorrect result,
\begin{equation}\label{var}
{\rm Var} (G) =\frac{M^2}{4} {\rm Var}(\sigma) =
\frac{M^2}{4}\left(\frac{T}{F+M-1}\right)^2 \rightarrow
\frac{T^2}{4},
\end{equation}
where the last expression is obtained for $M \gg 1$. Thus, for the
perfect coupling, we would have ${\rm Var} (G) =1/4$, which is twice
the correct value.

In order to reveal the role of correlations between different cross
sections, instead of Eq. (\ref{var}) we write the relation that
takes into account all the correlations,
\begin{equation}\label{var+corr}
{\rm Var} (G) = \frac{M^2}{4}\left(\frac{T}{F+M-1}\right)^2 + N^\ast
C_\Sigma + (N_c -N ^\ast) C_\Lambda ,
\end{equation}
where $N_c=L(L-1), \,N^\ast=L(M-2),\, L=M^2/4$. Here the terms
$C_\Lambda$ and $C_\Sigma$ stand for the correlation functions,
\begin{equation}\label{lamb}
C = \overline{\langle \sigma^{ab} \sigma^{a^\prime b^\prime}
\rangle} -\overline{\langle\sigma^{ab}\rangle
\langle\sigma^{a^\prime b^\prime}\rangle}
\end{equation}
with one common index (either $a=a^\prime$ or $b=b^\prime$) for the
$\Sigma-$correlations, and with no common indexes (both $a\neq
a^\prime$ and $b\neq b^\prime$) for the $\Lambda-$correlations. The
bar in Eq. (\ref{lamb}) represents the average over different
possible combinations of $a,b,a^\prime,b^\prime$, and
$\langle...\rangle$ indicate the average over energy. Our analysis
shows that for $M \gg 1$ one obtains, $C_\Sigma \approx -M^{-3}$ and
$C_\Lambda \approx 2 M^{-4}$. Therefore, in the limit of large
number of channels, one can write, in the correspondence with the
structure of Eq.(\ref{var+corr}),
\begin{equation}\label{1-8}
{\rm Var} (G) = \frac{1}{4}-\frac{1}{4}+\frac{1}{8}=\frac{1}{8}.
\end{equation}
This result clearly demonstrates the crucial role of correlations
for the variance of the conductance. These correlations are
neglected in the traditional theory of Ericson fluctuations.

\begin{figure}[h!]
\centering
\includegraphics[width=4.5cm,angle=-90,viewport=150 360 480 440]{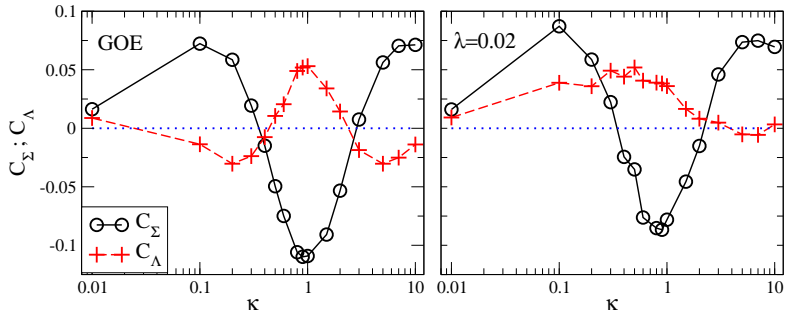}
\caption{Dependence of average correlations between cross sections
on the coupling $\kappa$ for the GOE case (left panel) and for a
weak interaction with $\lambda=0.02$ (right panel). Circles stand
for the $C_{\Sigma}$ and for the $C_{\Lambda}$ correlations (refer
to text for details). Both panels are on the same scale for $n=6$
and $m=12$. }
\end{figure}

Of special interest is the $\kappa-$dependence of the correlations
for a non-perfect coupling, in the regions $\kappa < 1$ and $\kappa
> 1$. As one can see, the $\Lambda-$ and $\Sigma-$correlations have
very different behavior, both for weak and strong intrinsic chaos.
The remarkable fact is that, for the perfect coupling with
$\kappa=0$, the $\Lambda-$ and $\Sigma-$correlations have maximal
values of opposite signs. These results may be important for the
analysis of experimental nuclear or solid state data.

\subsection{Conclusions}

In conclusion, we have studied the statistics of resonance widths
and cross sections for a fermion system coupled to open decay
channels. We used the effective non-Hermitian Hamiltonian that
reflects exact quantum-mechanical features of many-body dynamics in
an open interacting system. Main attention was paid to various
signatures of the crossover from isolated to overlapping resonances
in dependence on the strength of inter-particle interaction modeled
by the two-body random matrices in the intrinsic part of the
non-Hermitian Hamiltonian.

In the framework of our model we have tested main predictions of the
Ericson fluctuation theory. We have shown that the assumption that
the fluctuations of the resonance widths become negligible for a
large number of channels, is invalid in the overlapping regime. We
found that the fluctuations of resonance widths increase along with
the coupling to the continuum, resulting in the divergence of the
relative fluctuation of the widths (the ratio of the variance to the
square of the average width) at the transition point $\kappa=1$ for
any number of channels.

We have also shown that, in agreement with previous studies, the
correlation length differs from the average width for any number of
channels. On the other hand, the Weisskopf relation (\ref{62}) that
connects the correlation length of the cross section to the
transmission coefficient, works, for a large number of channels, at
any value of the intrinsic interaction strength $\lambda$. In many
situations the data show that increase of $\lambda$ suppresses the
fluctuations in the continuum. This can be understood qualitatively
as a manifestation of many-body chaos that makes all internal states
uniformly mixed \cite{big}.

Our results can be applied to any many-fermion system coupled to the
continuum of open decay channels. The natural applications first of
all should cover neutron resonances in nuclei, where rich
statistical material was accumulated but the transitional region
from isolated to overlapped resonances was not studied in detail.
Other open mesoscopic systems, for example, quantum dots and quantum
wires, should be analyzed as well in the crossover region.

It was our special interest to study the mesoscopic fluctuations of
the conductance in the dependence on the coupling strength and
degree of intrinsic chaos. In particular, we have analyzed how the
correlations between different cross sections influence the variance
of the conductance fluctuations. Our data manifest that these
correlations determine the universal value 1/8 (see, for example,
\cite{cond-fluct,been}) for the conductance variance in the perfect
coupling regime, $\kappa=1$, and for very strong interaction between
particles. Thus, we demonstrate that the Ericson theory of
cross-section fluctuation that neglects these correlations, is
insufficient for the description of the mesoscopic fluctuations.

\begin{acknowledgments}
We acknowledge useful discussion with T.~Gorin, Y.~Fyodorov,
D.~Savin, and V.~Sokolov. The work was supported by the NSF grants
PHY-0244453 and PHY-0555366. The work by G.P.B. was carried out
under the auspices of the National Nuclear Security Administration
of the U.S. Department of Energy at Los Alamos National Laboratory
under Contract No. DE-AC52-06NA25396. S.S. acknowledges the
financial support from the Leverhulme Trust.
\end{acknowledgments}

%%%%%%%%%%%%%%%%%%%%%%%%%%%%%%%%%%%%%%%%%%%

\end{document}